\documentclass[showpacs,prl,twocolumn]{revtex4}

\usepackage{epsfig}
\usepackage{amsmath}
\usepackage{graphicx}
\usepackage{dcolumn}
\usepackage{amsmath}
\usepackage{latexsym}

\begin{document}

\title{Coexistence of Coulomb blockade and zero bias anomaly in a strongly
coupled quantum dot}
\author{L. Bitton}
\author{D.B. Gutman}
\author{R. Berkovits}
\author{A. Frydman}
\address{The Department of Physics, Bar Ilan University, Ramat Gan 52900,
Israel}

\begin{abstract}
The current-voltage characteristics through a metallic quantum dot which is
well coupled to a metallic lead are measured.  It is shown that the I-V
curves are composed of two contributions. One is a suppression of the
tunneling conductivity at the Fermi level and the second is an oscillating
feature which shifts with gate voltage.  The results indicate that
Zero-Bias-Anomaly and Coulomb Blockade phenomena coexist
in an asymmetric strongly coupled quantum dot.
\end{abstract}


\date{\today}

\maketitle
Electron-electron interactions have a dramatic influence on the electronic
properties of low dimensional systems. Since the dawn of solid state physics
it is customary to identify two distinct contribution of interactions to
electronic properties (such as the electron self energy) in solids:
the Hartree and exchange terms.
While the Hartree term represents the contribution of the classical
Coulomb interaction, the exchange term correspond to much subtler
quantum contributions.

In the tunneling conductance these two terms manifest themselves in
two distinct ways: The suppression of the tunneling density of states at
the Fermi level, known as the zero bias anomaly (ZBA),
and the Coulomb gap/blockade. For example, tunneling into low-dimensional weakly disordered metals
exhibit a pronounced ZBA phenomenon in two dimensional disordered
metallic  films \cite{AA2} and 1D wires \cite{Glazman}, while
the Coulomb gap phenomenon dominates strongly disordered systems \cite{ES}.
Experimentally, both the ZBA and Coulomb gap manifest themselves as a sharp dip in the tunneling
conductivity at low  source-drain voltage, $V_{SD}$, though the two effects have a different functional form.
Since both features are centered on the Fermi energy ($V_{SD}=0$) it is very hard
to separate them.

Both terms influence also the tunneling into a quantum dot which
is coupled to leads. For strongly coupled dots one expects the
exchange term to dominate, resulting in a ZBA. For
weakly coupled dots, the Hartree term prevents tunneling conductivity except at the degeneracy point.
This leads to a zero conductance plateau in the low V regime of the I-V
characteristic known as the Coulomb blockade, CB. In the case of an asymmetrical quantum dot, which
is coupled more strongly to one of the leads than the other, the I-V
curve exhibits a series of differential-conductance plateaus termed
the Coulomb staircase (CS). Unlike the ZBA feature in which the
conductance minimum is pinned to the Fermi level, the CS
is sensitive to the chemical potential of the dot and oscillates with
the gate voltage $V_{g}$ \cite{CS}. This provides a clear way to
distinguish between CB and ZBA experimentally.

For most coupling strength one of these effects will dominate over the other. For a weakly coupled dot  ($g \ll 1$ where $g = h G / e^2$, $G$ corresponding
to the conductance of the dot-lead system), Coulomb blockade completely suppresses tunneling at small voltages, thus overshadowing any other e-e contribution. For
tunneling into an asymmetrical strongly coupled dot ($g \gg 1$)
the Coulomb blockade vanishes and the I-V curve should exhibit only a ZBA feature.
What happens between these two regions, i.e., for $g \geq 1$?
Can both effects be separated in the intermediate coupling regime?
In this Letter we shall experimentally address these questions.
Since the CB amplitude is predicted to decrease exponentially with $g$
\cite{wang,nazarov}, there can not be a wide coupling regime in which CB
has not yet completely vanished while it is suppressed enough to allow
a measurable ZBA and the two phenomena can coexist. A theoretical
answer to these questions is given in a paper by
Golubev {\it et al.} \cite{golubev} who considered the I-V characteristics
of an open quantum dot characterized by resistances $R_{S}$ and $R_{D}$
between the dot and the source/drain  electrodes respectively.
The electric current through the dot was found to be \cite{footnote}:

\begin{eqnarray}
\label{eq_golubev}
I(V)=G_{as}V-I_0(T,V)-\tilde{G}e^{-F(T,V)}V\cos 2\pi N\,.
\end{eqnarray}

Here the first term describes an  Ohmic current, characterized by linear
conductance  $G_{as}=1/(R_S+R_D)$;
the second term  [$I_0(T,V)$] reflects a conductance dip
at $V=0$ which we argue corresponds to
a ZBA. The last term  in  Eq. (\ref{eq_golubev})
describes a residual CB, i.e. a part of the current that
periodically oscillates with average number of electrons in the dot
\begin{equation}
N=\frac{C_SR_S-C_DR_D}{e(R_S+R_D)}V_{SD}+\frac{C_g}{e}V_g\,.
\end{equation}

In this Letter we present I-V characteristics of a dot which is strongly
coupled to a lead. By applying a unique method we are able to control this coupling in the regime of interest ($g \sim O(1)$). We find that the curves are composed of two parts,
corresponding to two of the terms in Eq. (\ref{eq_golubev})
one which is oscillatory with the source-drain voltage, $V_{SD}$, and shifts
with the application of gate voltage and the other, a conductance dip
centered at  $V_{SD}=0$ which is $V_g$ independent. The oscillatory feature
is dramatically suppressed as the dot is increasingly coupled to the lead
or as the temperature is raised. These results are interpreted as a
superposition of electron interaction induced DOS suppression and the classic CS simultaneously present
in a strongly coupled 0D system such that both ZBA and CB contribute to the same measurement.

\begin{figure}
{\epsfxsize=4 in \epsffile{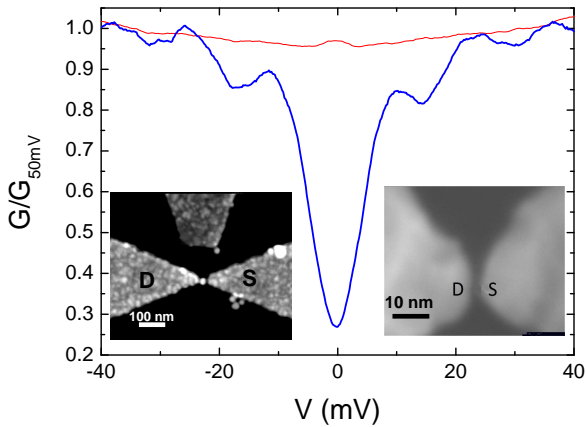}}\vspace{-1.3cm}\caption{Normalized
differential conductance as a function of source drain voltage for tunneling
directly between two gold electrodes (red light solid line) and tunneling
through a gold particle (heavy solid blue line). In both cases the tunneling
resistance at large voltage is $1.2 M\Omega$. T=4.2K. The insets show
scanning tunneling microscope images of both samples. Left inset: A 30nm
gold particle strongly coupled to the drain (left) electrode and weakly
coupled to the source (right) electrode. A gate electrode is fabricated at
a distance of 150nm. Right inset: Two gold electrodes placed a few nm apart
allowing direct source-drain tunneling.  \vspace{-0.5cm} \small}
\end{figure}

The quantum dots used in this work were Au nanoparticles, 30nm in diameter.
Coupling to a set of leads is performed in the following
way \cite{liora1, liora2}: On a Si-SiO substrate we fabricate two gold
electrodes (source and drain) separated by a gap of 10-30nm and a
perpendicular side gate electrode at a distance of 150nm as shown in the
left inset of Fig. 1. We then deposit an adhesive layer of Poly-L-Lysine
on the substrate and spread gold colloids on top. Next, we use Atomic Force
Microscope (AFM) nano-manipulation to "push" a desired colloid to the gap
between the source and drain electrodes. This yields a quantum dot which
is better connected to one of the leads than the other. We vary the coupling
of the dot to the lead by depositing gold atoms on top of the electrode
using an electrodeposition method. The substrate is placed in a solution
containing potassium cyanaurate, potassium bicarbonate and potassium hydroxide
\cite{marcus}. A deposition current of $1\mu A$ is applied between the drain
electrode and a gold counter electrode placed in the solution. This results
in an extremely fine controllable atomic gold growth on the electrode.
For each stage of the growth we cool the system to T=4.2K and measure I-V
characteristics at different gate voltages.

Since the dot is better coupled to the left electrode (the drain), one
can expect $R_{D} \ll R_{S}$, $R_{D}$ and $R_{S}$ corresponding to the
resistance between the dot and the drain and the source electrodes
respectively. Thus, the measured resistance through the dot $R=1.2M\Omega$
reflects $R_{S}$. On the other hand, both the ZBA and CB effects
are governed by the lower resistance, $R_{D}$, which is not directly
measurable by the conductance.

Fig. 1 depicts the tunneling conductance between the source and the drain
in two different cases: in the presence and absence of an Au dot connected
to the drain. It is seen that in the absence of a dot, the tunneling
conductance is nearly ohmic with a very small ZBA signature. This is
expected since the electrodes are relatively clean (resulting in large
conductance).  When a dot is introduced and strongly coupled
to the drain, the tunneling curves change drastically as seen in the
heavy solid line of Fig. 1. A large conductance minimum centered around
$V_{SD}=0$ is observed accompanied by a series of conductance oscillations.
We have observed similar conductance versus voltage curves for over
ten similar samples, all yielding very dramatic conductance minima accompanied by a superimposed oscillatory feature.

\begin{figure}
{\epsfxsize=3.8 in \epsffile{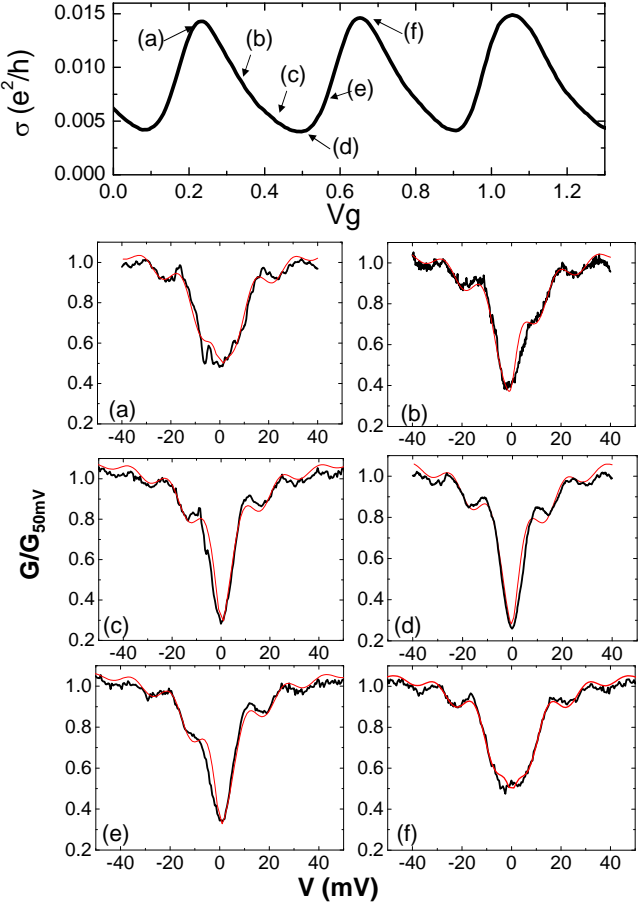}}\vspace{-1cm}
\caption{Top: conductance versus gate voltage at low $V_{SD}$ of a quantum dot system
measured at T=4.2K exhibiting CB features. a-f: Differential conductance
versus  $V_{SD}$ for a series of gate voltages corresponding to the marked
points in the top figure. The light solid red lines are fits
to Eq. (\ref{eq_fit}). \vspace{-0.7cm}
\small}
\end{figure}

Applying gate voltage, $V_g$, modifies these traces in a non-trivial way.
The conductance versus gate voltage at low $V_{SD}$ depicted in the top panel of Fig. 2
reveals pronounced conductance oscillations, which are attributed to the CB
phenomena \cite{liora2}. Fig. 2 shows a series of differential conductance
versus $V_{SD}$ curves taken for different gate voltages which span a
typical CB oscillation.

The analysis of our findings is based on the theoretical treatment of
a strongly coupled dot by
Golubev {\it et. al.} \cite{golubev} summarized in Eq. (\ref{eq_golubev}).
The general expressions for the functions  $I_0(T,V)$ and $F(T,V)$
are quite cumbersome and not very transparent \cite{golubev}.
However they can be considerably  simplified for the experimentally
relevant situation. For the low temperature limit  ($T \ll \hbar/2\pi\ R_D C$)
and asymmetrically coupled dot ($R_S \gg R_D$)  one  finds

\begin{equation}
I_0(T,V)=\frac{e^2}{2\pi\hbar}\frac{R_D}{R_S}V
\log\left(1+\frac{\hbar^2}{4 \pi^2 \epsilon^2 t_c^2}\right)\,,
\end{equation}

where  $\epsilon={\rm max}\{eV,T\}$ and  $t_c=R_D C$.

For the exponential factor $F(T,V)$ one finds

\begin{eqnarray}
F(T,V)=\frac{\hbar}{2e^2 R_D}+\frac{2\pi^2}{e^2}CT+\frac{2\pi}{e^2}
\sum_{r=D,S}\frac{1}{R_r}y(x_r).
\end{eqnarray}

Here $x_r=R_SR_DR_r eVC/(R_S+R_D)^2$, and
$y(x)= x\arctan x  -\frac{1}{2}\ln\left(1+x^2\right)$
and $C=C_S+C_D+Cg$.

Differentiating Eq. (\ref{eq_golubev}) utilizing the expressions derived
in Eqs. (3) and (4)  and assuming a very asymmetric dot at small voltages
we reach the following expression for the differential conductance
through the dot:

\begin{eqnarray}&&
\label{eq_fit}
\frac{G(V)}{G_{as}}=
1-\frac{1}{g_{D}}\left(\ln\left[1+
\frac{\hbar^2}{(2\pi t_{c}\epsilon)^2}\right] -\frac{2}{1+\left(2\pi t_c\epsilon/\hbar \right)^2}\right)\nonumber \\&&
\!+\!\exp\left[-\frac{g_{D}}{2}-\frac{2\pi^2}{\tilde{E_C}}
\left(\!T\!+\!\frac{R_D}{R_S}V_{SD}\right)\right]
\!\cos\left(\!\frac{2\pi V_{SD}}{E_{C}}\!+\!\phi\right) ,
\end{eqnarray}

where $g_{D}=\frac{h}{e^2 R_D}$ is the dimensionless conductance
between the dot and the
drain (the well connected electrode), $t_{c}$ is the charging time
of the dot, $E_{C}=\frac{e^2}{C_{S}}$ is the charging energy determining the staircase
period, $\tilde{E_C}=\frac{e^2}{C}$ determines the amplitude of the CB oscillations, $\phi=CgVg/e$ is the phase of the CS which is sensitive to the
applied gate voltage and $\epsilon=(V^2+\tilde{T}^2)^{1/2}$ is the energy of the system. To  fit our data we  notice that the smearing of ZBA
is determined by  an effective temperature $\tilde{T}=2mV$,
which is a factor of 5 larger than the experimental temperature and will be discussed later on.

\begin{figure}
{\epsfxsize=3.8 in \epsffile{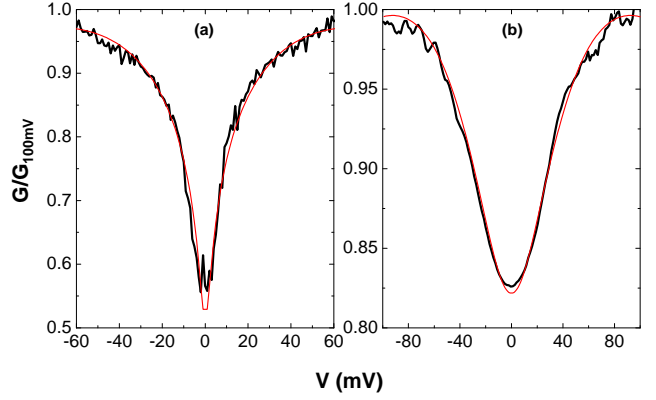}}\vspace{-1cm}\caption{
(a) Conductance versus SD voltage for the case where the dot of Fig. 2 was
coupled more strongly to the drain. The light solid red line is the fit to
Eq. (\ref{eq_fit}). (b) Conductance versus SD voltage for the dot of Fig 2.
at $T=77k$ with the fit to Eq. (\ref{eq_fit}). In both cases only a
dip around $V_{SD}=0$ appears with no measurable superimposed oscillatory
feature. \vspace{-0.5cm} \small}
\end{figure}

Eq. (\ref{eq_fit}) includes two distinct parts. The first term is a
conductance dip centered at $V_{SD}=0$ which we interpret as a
0D version of the ZBA for the well connected dot. As for the higher
dimensional cases discussed by Altshuler and Aronov\cite{AA2},
the dip magnitude is inversely proportional
to $g$ and always pinned to the Fermi energy, i.e., centered at $V_{SD}=0$.
The second term is oscillatory with $V_{SD}$, with a period corresponding
to the charging energy of the dot $E_C$. This term corresponds to the
CB, it is sensitive to the gate voltage via the phase $\phi$ and is suppressed exponentially with $g_D$ as predicted
for the CB phenomena \cite{wang,nazarov} and with the $V_{SD}$ and T. We emphasize that the same fitting parameters were used for all cases with
only $\phi$ varying between the different curves a-f, thus "sliding"
the CS along the voltage axis. These fits yield $g_{D}=3.5$
or $R_{D}=7.4k\Omega$, demonstrating that, contrary to the orthodox
convention, CB effects can be measured even for $g>1$. We note that
we use the same $g_{D}$ for both terms of Eq. (\ref{eq_fit})
thus increasing our confidence in the fitting procedure.

The CB term shows a difference of its suppression by temperature
and source-drain voltage. While the temperature scale is determined
by $\tilde{E_C}$, the voltage suppression is determined by $\frac{R_S}{R_D} \tilde{E_C}
\gg \tilde{E_C}$. This explains why CB oscillations can be seen for $V_{SD} \sim
50 mV$ while for $T = 77K \sim 8 mV$ they are completely suppressed as seen in Fig. 3.
The physical origin of the different behavior of the temperature and
SD voltage stems from their different influence on inelastic processes
of electrons in the dot. While
temperature affects the occupation of all the
electrons in the dot, the SD voltage influences only electrons
tunneling in or out of it. Since the vast majority of electrons
enter (or leave) the dot through the low resistance connection
to the lead the relevant voltage scale is proportional to the voltage
drop on it, i.e., to
$\frac{R_S}{R_D} V_{SD}$.

As the coupling between the dot and the drain is increased all non-ohmic
features in the I-V are suppressed. However, the oscillatory feature is
suppressed much faster. Fig. 3a shows the differential
I-V curve for the
same dot depicted in Fig. 2 in which the coupling to the drain has been
increased. The I-V in this case exhibits only a ZBA like feature with no
signs for CS effects. For this coupling the fit to
Eq. (\ref{eq_fit}) is achieved for $g_D=5.5$. This result indicates
that the CB effect is much more sensitive to coupling than the ZBA feature,
as expected from Eq. (\ref{eq_fit}).

Extracting $g_D$ out of the ZBA feature in $11$ measurements
performed for different dots and comparing the amplitude of the CB
oscillations of each case at $V_{SD}\approx0$
(see Fig. 4) one can clearly see an exponential decrease of the CB amplitude as $g_D$ increases which fits very well the explicit expression appearing in Eq. (\ref{eq_fit}),
i.e. the amplitude decreases as $\exp(-g_D)$. This fit provides an experimental verification of the common theoretical prediction that
the CB amplitude should be suppressed exponentially with g in the regime $g\geq 1$ and also reinforces the confidence in the consistency of the analysis.

\begin{figure}[tb]
{\epsfxsize=3.4 in \epsffile{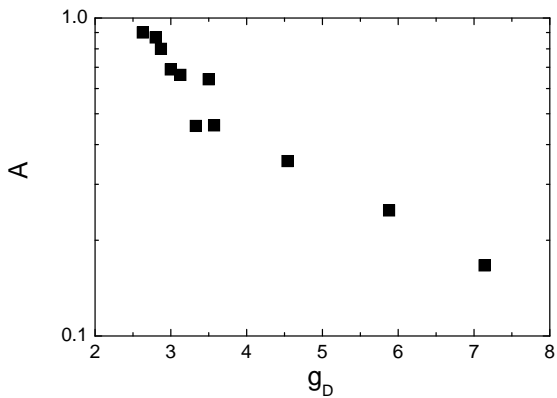}}\vspace{-1cm}\caption{
Amplitude of the CB oscillations at small $V_{SD}$ (such as those of Fig. 2a) defined as $[g_{max}-g_{min}]/g_{max}$ as a function of $g_{D}$ extracted from the fits to Eq. (\ref{eq_fit}) for 11 different dots. \vspace{-0.5cm} \small}
\end{figure}

A similar suppression is seen when the temperature is increased. Fig. 3b shows
the I-V characteristics through the dot at $T=77K$. It is seen that at this
temperature only a ZBA-like feature is observed and the CS has
vanished. Indeed, CB effects are expected to decay exponentially with temperature (see Eq. (\ref{eq_fit})). Nevertheless an effective temperature of
$\tilde{T}=27mV$ in the ZBA term (more than 4 times the actual temperature)
is needed to achieve a good fit (light solid red line).

Thus the experimental results seem to be well described by
our approximation of the
theory of Golubev {\it et. al.} \cite{golubev} depicted in Eq. (\ref{eq_fit}).
There remains though the puzzle of the relative high effective
temperature $\tilde T$ needed for the fit of the ZBA term in
Eq. (\ref{eq_fit}). $\tilde T$ corresponds to
smearing of the ZBA dip. The fact that the smearing of the dip
is stronger than expected from the system's temperature indicates
that an additional physical mechanism contributes to inelastic
processes in the dot. In contrary to typical quantum dot systems
the grain here is mechanically free standing, although there is a good
electrical contact. Mechanical vibrations, which might induce
fluctuations in the tunneling to the dot might significantly
contribute to the dephasing and inelastic processes manifested
in the smearing of the ZBA.

In conclusion, we have investigated the I-V characteristics of transport
through an asymmetrically coupled metallic grain, which is well connected
to one of the leads, but poorly connected to the other. This is an
unusual regime for which properties usually associated with weakly coupled
0D systems (CB staircases) and disordered higher dimensional
leads (Altshuler-Ahronov ZBA) appear and
may be easily separated. Identifying the relevant
limits in the general expression of Golubev {\it et. al.} we have shown
that in our case the CB and ZBA are both relevant and show a different
dependence on the strong coupling of the grain to the lead ($g_D$),
temperature, gate and source-drain voltage. We use these features
to determine the parameters of the transport (such as the source and drain
resistances) which can not be separated in regular transport measurements.

We are grateful for useful discussions with Y. Gefen and A.D. Mirlin  and especially insightful input from I.V. Gornyi. This research was supported by the Israeli academy of science (grant number 399/09)


\begin{references}
\bibitem{footnote} Some  of numerical coefficients in
Eq.(\ref{eq_golubev}) are not universal and  depend on
transmission amplitudes $t_n$ between the dot and the
leads\cite{golubev2,Yeati}.
Since the  values of $t_n$ are neither known nor expected
to strongly influence  our  results  we use the simplest model of
$t_n \ll 1$.
\end{references}
\end{document}